\documentclass[sigconf,screen,nonacm]{acmart}

\usepackage{booktabs}
\usepackage{graphicx}
\usepackage{microtype}
\usepackage{balance}
\setcopyright{none} 
\title{Recompilation Is Not Enough: Test-Guided Decompiled-C Repair}

\author{Yuhan Huang}
\affiliation{%
  \institution{Xidian University}
  \city{Xi'an}
  \country{China}
}
\affiliation{%
  \institution{Ant Group}
  \city{Hangzhou}
  \country{China}
}

\author{Puzhuo Liu}
\affiliation{%
  \institution{Ant Group}
  \city{Hangzhou}
  \country{China}
}

\author{Jianlei Chi}
\affiliation{%
  \institution{Xidian University}
  \city{Xi'an}
  \country{China}
}

\renewcommand{\shortauthors}{Yuhan Huang, Puzhuo Liu, and Jianlei Chi}

\begin{abstract}
Decompiled C often becomes recompilable only after repair, but recompilation alone does not establish test-observed behavior. A recompiled command-line binary can still parse options incorrectly, print different bytes, or return a different exit status. We present a few-step workflow for repairing decompiled C using compiler feedback and related official tests. Compiler and linker diagnostics first guide build repair. Once the repaired C recompiles into a binary, smoke checks and related official tests expose behavioral discrepancies for semantic repair. In a preliminary static-enriched evaluation on 104 Coreutils 9.5 binaries with available decompiler exports and deterministic exact-output smoke comparisons, 91 binaries (87.5\%) recompile and pass the test gate; 9 do not recompile within the repair budget, and 4 recompile but still fail the test gate. The result suggests that test-gate feedback can make LLM-assisted repair of decompiled C more auditable than compile-only recovery.

\end{abstract}


\begin{document}

\maketitle

\section{Introduction}

Decompilers can generate seemingly plausible C code for reverse engineering or taint analysis\cite{liu2025llm,liu2023fits}. 
The decompiled result can even be recompiled, but the resulting binary may still parse options differently, print different bytes, or return different exit statuses.
This paper studies that gap for complete command-line binaries. Recent work has made LLM-based binary decompilation more concrete: assembly-language models and large decompilation benchmarks broaden the training and evaluation setting~\cite{jiang2025nova,tan2025decompilebench}, two-phase decompilation separates structural recovery from identifier recovery~\cite{tan2026sk2decompile}, and refinement systems target recompilable output, distorted pseudocode, or decompilation fidelity~\cite{wong2025decllm,li2025pseudofix,zhou2026fidelitygpt}. For decompiled C, the remaining repair gap is behavioral: a recompiled binary can still diverge from the original utility's command-line behavior.

The gap is sharper when the repair unit is a complete binary rather than an individual function. Function-level decompilation settings can evaluate recovered functions in controlled harnesses~\cite{armengol2024slade,tan2025decompilebench}, but a repaired Coreutils binary must recover option tables, static data, global state, helper-library behavior, output formatting, and exit-status conventions. Recent evaluation work separates readability, recompilability, and functionality~\cite{liu2026debench}, while equivalence checking, fidelity studies, type recovery, and broader binary-analysis benchmarks show that recovered C and binary-analysis outputs can fail along different axes~\cite{dramko2025codealign,dramko2024taxonomy,wang2025typeforge,shang2025binmetric}. We therefore treat recompilation as an intermediate step and use related official tests as repair feedback.

The workflow repairs one decompiled Coreutils binary at a time. It first uses compiler and linker diagnostics to repair the decompiled C until it recompiles. It then validates behavior with smoke checks and related official tests. We use related official tests to mean Coreutils tests drawn from the official suite and assigned before repair because they exercise the target binary's expected command-line behavior. They are not Coreutils source files or complete answer files; compact failure summaries may include observable discrepancy fragments. Test failures are summarized into repair evidence, and the next repair step uses those discrepancies to modify the decompiled C. We do not use Coreutils source code as repair input.

This setting differs from both decompiler readability evaluation and source-level automated repair. Decompiled C may contain type reconstruction errors, missing global data, synthetic control flow, and library-boundary confusion at the same time. The repair target is also not a localized regression in a maintained program; it is a recompiled binary whose observable command-line behavior should match the original utility under the test gate. For that reason, the workflow is deliberately conservative: a binary is accepted only if the repaired C recompiles and the recompiled binary passes related official tests and smoke checks.

This work-in-progress paper makes two contributions: (1) a few-step workflow for repairing decompiled C at binary granularity, and (2) a test-gate feedback loop that guides semantic repair beyond recompilation.

As preliminary evidence, we evaluate 104 GNU Coreutils 9.5 binaries in the static-enriched track, restricted to binaries with available decompiler exports and deterministic exact-output smoke comparisons. Under a bounded few-step policy with a primary three-iteration build target, recorded diagnostic extensions, and a three-iteration semantic cap, 91 of the 104 binaries (87.5\%) recompile and pass the test gate; 9 do not recompile within the repair budget, and 4 recompile but still fail the test gate.

\section{Motivation}

Recompilation exposes build-level faults, but it does not establish behavior. Compiler diagnostics identify missing declarations, type conflicts, and unresolved symbols; they rarely explain why an option combination should produce a specific byte sequence or exit status. Smoke checks are useful sanity checks, but related official tests can expose boundary cases, option interactions, and user-visible conventions accumulated by the utility maintainers.

The distinction also determines which repair evidence is useful. A compiler error usually points to a local syntactic defect, while a behavioral failure often reflects a cross-function invariant: a global option table may be malformed, a recovered helper may have the wrong behavior, or a static data object may be present but laid out incorrectly. Treating all failures as generic LLM prompts encourages trial-and-error. Treating them as staged evidence lets the workflow ask first whether the failure is a recurring decompiler artifact, and only then whether a semantic edit is needed.

Related official tests also constrain the repair target better than ad hoc examples. A single hand-written smoke command can show that a binary starts and handles a common path, but it rarely exercises long-option aliases, diagnostic formatting, ordering rules, or error exits. Related official tests are not a proof of equivalence, yet they provide a stronger behavioral check than recompilation alone and a more reusable target than utility-specific manual scripts.

\begin{figure}[!t]
\centering
\includegraphics[width=0.8\columnwidth]{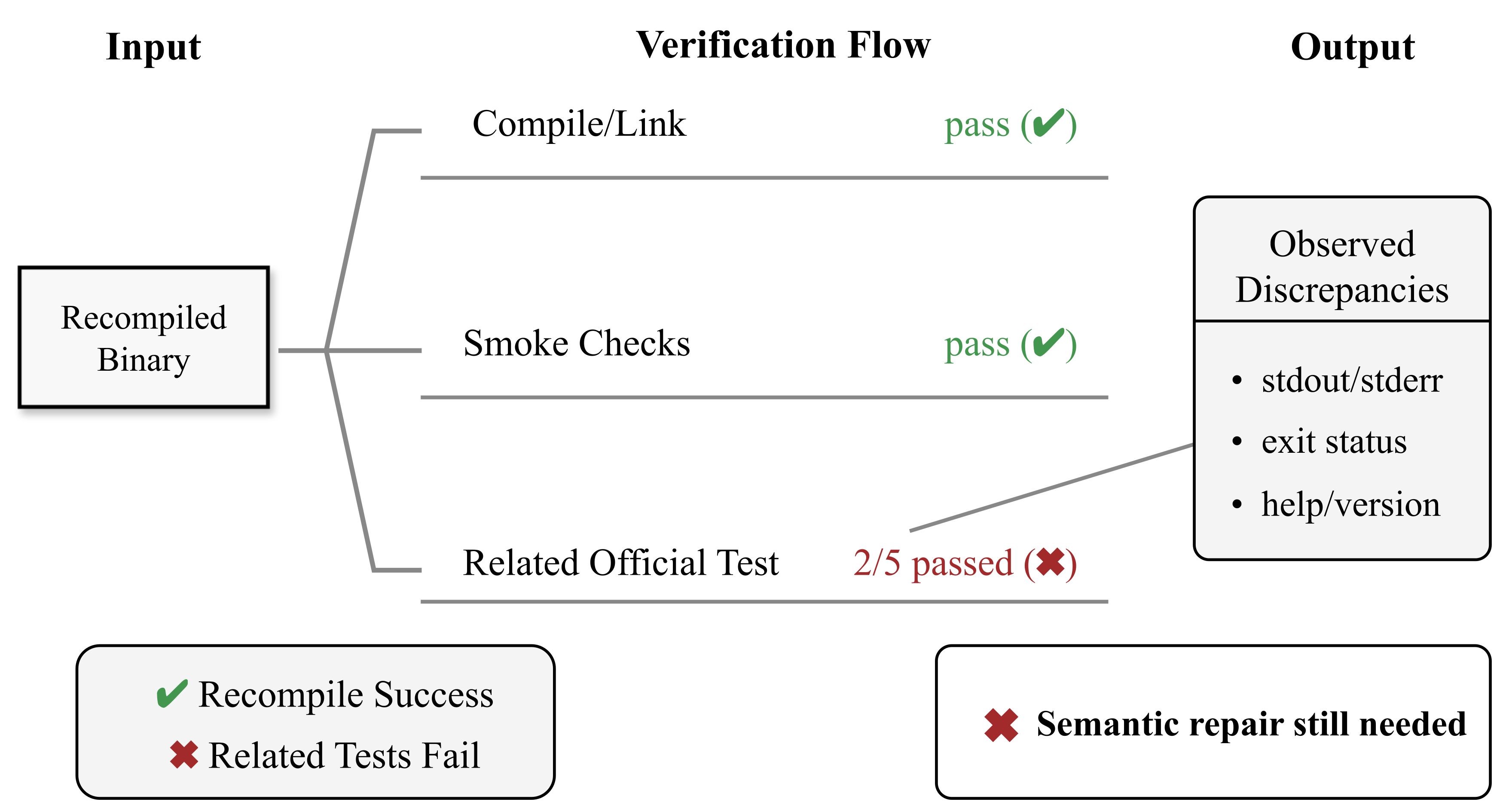}
\caption{Evidence gap in a \texttt{cut} repair case. Compile/link and smoke checks pass, but related official tests still expose test-observed behavioral discrepancies.}
\Description{Diagram showing a recompiled cut binary checked by compile/link, smoke checks, and related official tests. Compile/link and smoke checks pass, related official tests pass only 2 of 5 tests, and observed discrepancies include stdout/stderr bytes, exit status, and help/version behavior.}
\label{fig:cut-motivation}
\end{figure}

Figure~\ref{fig:cut-motivation} shows the evidence gap. Compiler and linker diagnostics establish that the repaired C can produce a binary, and smoke checks establish that a shallow command-line path works. In the \texttt{cut} case, however, the recompiled binary passed only 2 of 5 related official tests, with output, exit-status, and help/version discrepancies. The case is small, but it illustrates the paper's central point: related official tests expose repair evidence that compiler diagnostics and smoke checks do not capture.

\section{Approach}

\begin{figure*}[!t]
\centering
\includegraphics[width=0.8\textwidth]{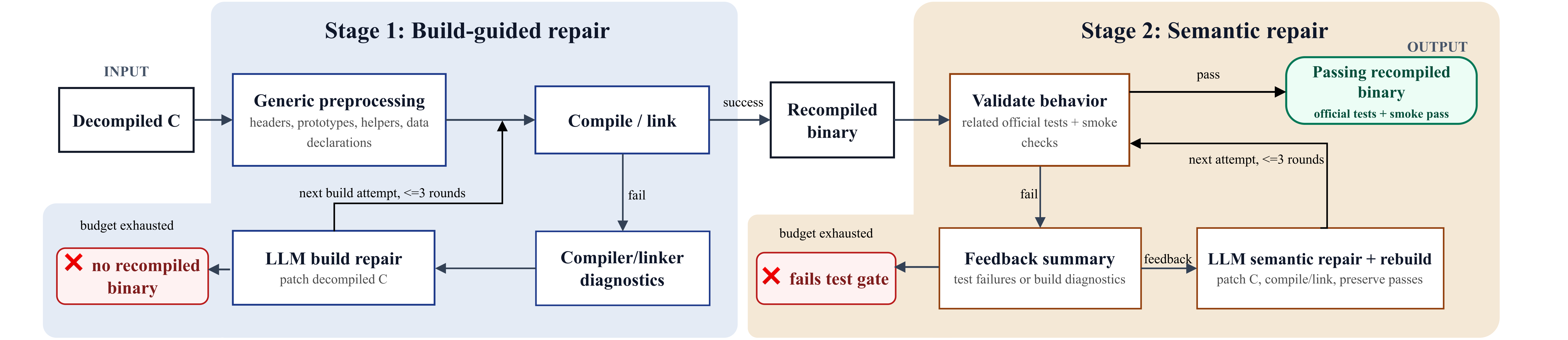}
\caption{Bounded two-stage repair workflow. Compiler/linker feedback guides build repair; related official tests and smoke checks guide semantic repair.}
\Description{Double-column workflow diagram. Decompiled C enters a build-guided repair stage with generic preprocessing, compile/link, compiler/linker diagnostics, and LLM build repair with a primary target of three build attempts and recorded diagnostic extensions where used; budget exhaustion reaches a no-recompiled-binary failure state. A recompiled binary then enters a semantic repair stage. Related official tests and smoke checks validate behavior. Passing those checks reaches a passing recompiled-binary state. Failing checks produce a feedback summary for LLM semantic repair and rebuild, which patches C, compiles and links it, preserves passes, and returns to behavior validation for up to three attempts; budget exhaustion reaches a fails-test-gate state.}
\label{fig:overview}
\end{figure*}

Figure~\ref{fig:overview} summarizes the workflow. The input is decompiled C for one Coreutils binary, optionally aided by sanitized static metadata such as recovered symbols, constants, and data-layout hints. The metadata comes from binary and static-analysis sidecars, not Coreutils source snippets or complete expected-output files. The output is either a passing recompiled binary or an explicit failure label.

The workflow has two feedback loops because build failures and behavioral failures have different causes. The first loop repairs the decompiled C until it recompiles. Its evidence is compiler and linker diagnostics. The second loop starts only after recompilation. Its evidence is behavior differences from smoke checks and related official tests; if a semantic edit breaks recompilation, the rebuild diagnostics are folded back into the same semantic-repair attempt. Keeping the loops separate avoids treating every test-gate failure as another compile error, or treating every compile error as if it required semantic reasoning. The separation also makes the final accounting auditable, because every failed binary has a stage label.

Every repair attempt records recompilation and test-gate status. Build repair has a primary three-iteration target with recorded diagnostic extensions, semantic repair is capped at three iterations, and failures remain explicit.

\textbf{Build repair.}
The first stage turns decompiler output into C that can be compiled and linked. Deterministic preprocessing handles recurring, source-independent artifacts: declarations and headers, decompiler intrinsic helpers, ABI and prototype cleanup, recovered global data declarations, and minimal C/POSIX helper shims. These fixes are justified by C/POSIX semantics, ABI conventions, binary-visible symbols, static metadata, or black-box original-binary behavior. If preprocessing does not produce a recompiled binary, LLM build repair receives the decompiled C and compiler/linker diagnostics. The primary build-repair target is three iterations; any bounded diagnostic extension is recorded as repair effort rather than hidden from the final accounting.

The deterministic rules are intentionally narrow. They normalize artifacts that appear repeatedly in decompiled C, such as inconsistent integer typedefs, missing libc prototypes, synthetic temporaries, or recovered data objects that lack declarations. A rule is acceptable only when its justification is independent of a particular test outcome. This keeps preprocessing closer to a reusable post-decompilation repair layer and reduces the risk that it becomes a hidden benchmark-specific solution.

\textbf{Test gate.}
Recompilation is not enough. After the repaired C recompiles, the workflow executes the recompiled binary and validates behavior with related official tests plus smoke checks. A binary passes the test gate only when it recompiles and passes both forms of behavioral evidence. Separately, generated repairs are audited against known test-specific branch patterns and hard-coded answer patterns. Unlike function-level decompilation settings, this gate executes the repaired binary as a standalone command-line program and exposes missing cross-function or global behavior~\cite{armengol2024slade,tan2025decompilebench}. Test evidence is assigned before repair, following the observation that repair models benefit from relevant facts but can degrade when supplied with indiscriminate context~\cite{parasaram2025factselection}.

Smoke checks and related official tests play different roles. Smoke checks are cheap and deterministic, so they catch severe regressions such as a binary that cannot start or mishandles a canonical command. Related official tests are more targeted and can expose boundary ranges, option interactions, diagnostic formatting, and error-path behavior that simple smoke inputs often miss. Requiring both checks prevents the workflow from accepting a binary that merely satisfies a shallow command-line trace.

\textbf{Semantic repair.}
When tests fail, the workflow summarizes what differed: failing tests, stdout/stderr mismatch summaries, exit-status mismatches, timeouts, and smoke-check results. Deterministic repair handles recurring defect classes such as malformed static data, recovered option tables, incorrect helper signatures, data-layout constants, and error-path conventions. These recoveries are broad categories, not complete answer files. Remaining cases are passed to LLM semantic repair with the compact discrepancy summary. Each semantic edit is rebuilt before retesting; if the edit introduces a compile or link failure, the resulting diagnostics are included in the next feedback summary for that semantic attempt. After three semantic-repair iterations, binaries that still fail the test gate are counted as failures. This bounded loop is related to recent LLM repair agents that iterate over execution and test feedback~\cite{yang2024sweagent,bouzenia2025repairagent}, but the starting point here is a compound decompiler artifact rather than a localized source-level bug.

The discrepancy summary is compact by design. It does not forward complete test logs or complete expected-output files as repair targets. Test identifiers and short discrepancy fragments may be retained for audit and context, while generated repairs are audited against known test-name branches, test-specific command-line branches, Coreutils source snippets, and answer-key branches. The summary emphasizes observable discrepancies and asks the repair step to modify the underlying utility logic while preserving already passing behavior. This is a practical compromise between giving too little evidence, which leads to blind edits, and giving unrestricted transcripts, which increases overfitting risk.

\section{Evaluation}

\subsection{Setup}
The preliminary evaluation uses 104 GNU Coreutils 9.5 binaries with available decompiler exports and deterministic exact-output smoke comparisons. All results use the static-enriched track: repair may use sanitized binary/static-analysis sidecars, including recovered symbols, constants, and data-layout hints, but excludes Coreutils source snippets and complete expected-output files.

The unit of accounting is the binary, not an individual function. A binary is accepted only when the repaired C recompiles, links, and the recompiled binary passes the test gate: related official tests plus smoke checks. Related official tests are Coreutils tests drawn from the official suite and assigned before repair because they exercise the target binary's expected command-line behavior. They provide behavioral evidence without providing Coreutils source files or complete answer files as repair input. Full Coreutils suite results, when collected, are stress/audit artifacts rather than the primary completion result.

Each binary is repaired under a bounded few-step policy: build repair has a primary three-iteration target with recorded diagnostic extensions, and semantic repair is capped at three iterations. Candidates that do not compile and link within the build budget are build failures; candidates that recompile but still fail the test gate after semantic repair are behavioral failures. The budget is part of the method because it keeps successes, extensions, and failures visible under the same accounting rule.

\subsection{Effectiveness}
Each binary is tracked as an end-to-end repair artifact with repair history, recompilation status, and test-gate status. This matters because repaired C alone can be misleading: it can recompile, pass smoke checks, and still fail related official tests. The evaluation therefore reports passing recompiled binaries and explicit failure stages for the rest.

\begin{table}[t]
\caption{Outcomes for 104 Coreutils 9.5 binaries under bounded repair.}
\label{tab:outcome}
\centering
\begin{tabular}{lr}
\toprule
Outcome under bounded repair & Count \\
\midrule
Pass test gate & 91 \\
Do not recompile within the repair budget & 9 \\
Recompile but still fail test gate & 4 \\
\midrule
Total evaluated & 104 \\
\bottomrule
\end{tabular}
\end{table}

Table~\ref{tab:outcome} reports the final outcome taxonomy. In the static-enriched track, 91 of the 104 evaluated binaries (87.5\%) recompile and pass the test gate. The 13 remaining cases are preserved as explicit failures under the same bounded budget: 9 do not recompile within the build-repair budget, while 4 recompile but still fail the test gate after semantic repair.

This taxonomy separates two failure modes that would be collapsed by a compile-only metric. The 9 build failures point to unresolved post-decompilation C integration problems, such as declarations, recovered data, cross-file linkage, or helper behavior needed to produce a standalone command-line binary. The 4 test-gate failures are recompiled binaries whose observable behavior still diverges under related official tests or smoke checks.

The \texttt{cut} case in Figure~\ref{fig:cut-motivation} illustrates the evidence gap that triggers semantic repair. The recompiled binary passed smoke checks, but related official tests exposed the remaining behavior gap. In the full repair trace, two semantic-repair iterations then produced a binary that passed the related official tests for this case with no smoke regression. The discrepancy summary captured observable mismatches that point to user-visible behavior, while the repair checks audited the edit against known test-name branches and other test-specific shortcut patterns.

The trace also shows why semantic repair is not just another build-repair pass. The related official tests covered ordinary data processing and command-line conventions, including range parsing, read/write error paths, and help/version handling. The first recompiled binary had enough structure to run, but the discrepancy summary exposed behavioral categories that compiler diagnostics could not explain.

\subsection{Efficiency}
Across the 104-binary scope, the logs record 6.87M tokens, computed by summing build-repair trace tokens and semantic-repair tokens across successes, build-budget failures, diagnostic extensions, and test-gate failures. Of 95 binaries entering the test gate, 91 passed and 4 failed after the bounded semantic loop; the other 9 are build-stage failures. These records profile prototype repair effort rather than model-independent runtime efficiency, since wall-clock time depends on runner load, test scheduling, and external service latency.

\section{Discussion}

The preliminary result supports a repair discipline rather than a final decompilation benchmark. Under one bounded budget, binaries that do not recompile remain build failures, and recompiled binaries that fail the test gate are not accepted on smoke checks alone. This accounting limits the extent to which unbounded retry or manual steering can obscure the source of progress.

The current evaluation remains narrow. It covers one utility family, related official tests rather than full equivalence, and one primary static-enriched track. The deterministic layer is evidence-backed, but its portability beyond Coreutils-style binaries remains an open question. The exact-output smoke checks also favor deterministic command-line behavior. A clean pseudo-only ablation, more detailed component ablations, broader utility families, and root-cause analysis by failure family are future work.

There are also measurement threats. Related official tests are stronger than smoke checks, but they are still a subset of full behavioral equivalence. Static metadata may make some repairs easier than a pseudo-only setting, so future experiments should isolate the value of metadata, deterministic normalization, LLM build repair, and LLM semantic repair. Finally, transferring the workflow to libraries, servers, or interactive programs will require different behavioral tests and different rules for defining an externally observable contract.

\section{Related Work}

\textbf{Recent LLM-based decompilation.}
Recent systems move beyond lexical similarity toward executability and richer binary context. Nova and Decompile-Bench strengthen assembly-language modeling and large-scale evaluation~\cite{jiang2025nova,tan2025decompilebench}, while SK2Decompile separates source-structure recovery from identifier recovery~\cite{tan2026sk2decompile}. PseudoFix targets distorted decompiled C pseudocode~\cite{li2025pseudofix}. Earlier formally published systems such as LLM4Decompile and SLaDe provide useful foundations for decompilation generation and optimized assembly decompilation~\cite{tan2024llm4decompile,armengol2024slade}; ReSym, SymGen, and GenNm show how LLMs can recover symbols and names from stripped binaries~\cite{xie2024resym,jiang2025symgen,xu2025gennm}. These systems improve recovered function quality, but they do not by themselves define when a complete recompiled binary should be accepted as passing a test gate.

\textbf{Repair and correction loops.}
Decompiler-output refinement increasingly uses recompilation, correction, or validation feedback. DecLLM~\cite{wong2025decllm} and FirmNamer~\cite{liu2025function} refine decompiler output for program understanding; FidelityGPT uses retrieval-augmented correction~\cite{zhou2026fidelitygpt}, and DeGPT optimizes readability~\cite{hu2024degpt}. These systems do not evaluate a Coreutils whole-binary repair workflow driven by related official tests plus smoke checks. Our binary-level setting must reconstruct cross-function and global behavior rather than assume function-level context.

\textbf{Evaluation and repair feedback.}
Liu et al.~\cite{liu2026debench} make the readability, recompilability, and functionality split explicit; Decompile-Bench broadens decompilation measurement, and BinMetric situates code understanding within a broader binary-analysis benchmark suite~\cite{tan2025decompilebench,shang2025binmetric}. Equivalence checking, type recovery, and decompiler fidelity studies identify why recovered C can remain semantically fragile~\cite{dramko2025codealign,wang2025typeforge,dramko2024taxonomy}. In program repair, fact selection and repair agents show that test and runtime evidence can guide LLM edits~\cite{parasaram2025factselection,bouzenia2025repairagent,yang2024sweagent}. This paper applies that feedback principle to decompiled C at binary granularity and uses related official tests not only as an endpoint but also as repair evidence.

\section{Conclusion}

Repairing decompiled C should not stop at recompilation. This paper presents a bounded workflow that uses compiler/linker diagnostics for build repair and test-gate feedback for semantic repair. On 104 Coreutils 9.5 binaries in the static-enriched track with available decompiler exports and deterministic exact-output smoke comparisons, 91 binaries recompile and pass the test gate. The result provides bounded repair evidence rather than a proof of equivalence: accepted binaries have recorded repair traces, and failures retain explicit stopping reasons. Future work will broaden benchmarks, add pseudo-only and component ablations, and improve root-cause analysis by failure family without relaxing the bounded repair policy.

\clearpage
\balance
\bibliographystyle{ACM-Reference-Format}
\bibliography{references}

\end{document}